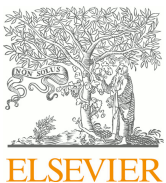
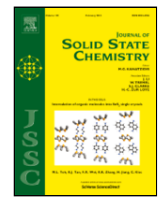

# Silica anchored colloidal suspension of magnetite nanorods

Nidhi Ruparelia, Urveshkumar Soni, Rucha P. Desai *, Arabinda Ray

*P D Patel Institute of Applied Sciences, Charotar University of Science and Technology, CHARUSAT Campus, Changa, Dist. Anand, Gujarat, 388421, India*

ARTICLE INFO

ABSTRACT

This study focuses on the interaction of colloidal silica nanoparticles with the magnetite ($Fe_3O_4$) magnetic fluids (MF), which eventually forms nanorod like structure. The aqueous magnetic fluid consists of magnetite nanoparticles having double layers of lauric acid surfactant. This surfactant provides stability towards short-range van der Waals attractive and steric repulsive forces, as well as long-range dipole-dipole interactive force. Whereas, in the colloidal silica, the sodium ions provide stability to the silica nanoparticles. The colloidal silica and magnetic fluid both were mixed in different proportions to understand the interaction between the silica and magnetite nanoparticles. Thus, the interaction present in the system is studied using FTIR, TGA, and magnetic field induced microscopy. The FTIR and TGA data reveal that silica interacts with the outer layer of lauric acid through the Si–O bond and eventually provides stability to the system. The length of the lauric acid sheath varies with the concentration of silica nanoparticles. The SEM images indicate nanorod formation, and its structure dimensions vary with the silica concentrations, which is also reflected in the magnetic field-induced structure formations. The structure observed using microscopy correlated with the interaction derived from FTIR and TGA data analysis. XRD data of iron oxide (magnetite) nanoparticles are present here again for completion.

## 1. Introduction

Interest in the stable colloidal suspension, either prepared using nanoparticles in the carrier liquid (e.g., hydrocarbon oil, water, solvent), or in the form of micro-emulsion (water in oil or oil in water), or by suspending both, micron and nano (magnetic/non-magnetic) material in polar/non-polar medium [1–4], increases owing to its tunable properties such as rheological, magnetic, optical, electrical, [5,6]. Silica is one of the widely used material, also known for its non-toxicity, adsorption properties, surface functionalization, thermal and chemical stability, and various biological and environmental applications [4,7,8].

The magnetic $Fe_3O_4$–$SiO_2$ core-shell nanorods have demonstrated applications for removal of water-soluble heavy metals like arsenic, lead (Pb(II)), and uranium (U(VI)) [9,10]. On the other side, researchers explored the magnetite needle-like nanorods and its core-shell architecture with silica for the direct methanol fuel cell (DMFC) preparations [9]. In this work, the chemical co-precipitation route was adopted to synthesize $Fe_3O_4$–$SiO_2$ nanorods. The core-shell structure was synthesized by silica (using Tetraethyl orthosilicate (TEOS)) shell on the core $Fe_3O_4$ spherical nanoparticles. Thus, nearly spherical prepared $Fe_3O_4$–$SiO_2$ particles were functionalized using an amine group followed by refluxing at 110 °C for 12 h under the nitrogen atmosphere. It resulted in final $Fe_3O_4$–$SiO_2$ core-shell nanorods. The process is lengthy with the usage of different solvent systems.

In comparison, Ranjani et al. synthesized $Fe_3O_4$ nanorods initially, which were coated by $SiO_2$ using TEOS as base chemical [11]. Khan et al. followed the chemical co-precipitation route to synthesize magnetite nanorods decorated by the Si-Schiff base complex [10]. In this, $Fe^{+2}$ and $Fe^{+3}$ salts were dispersed in ethanol, TEOS, and ammonia solution added and stirred for 24 h. In the finally obtained off-spherical $Fe_3O_4$–$SiO_2$ particles Si-Schiff based was added and stirred for 10 h, which results in rod-shaped $Fe_3O_4$/$SiO_2$-Si-SBC. Zhang et al. described the method to synthesized $Fe_3O_4$–$SiO_2$ nanoparticles and nanorods using a modified Stober method [12]. The magnetic colloidosomes consisting of $Fe_3O_4$–$SiO_2$ hetero-nanorods was prepared Using the water-in-oil emulsion.

Magnetic fluid (ferrofluid - a stable colloidal suspension of magnetic nanoparticles decorated by surfactant layer(s) suspended in magnetically passive medium) has attracted many researchers due to various promising engineering and biomedical applications [2,13–15]. The modulation of magnetic fluids properties (e.g., optical, rheological, electrical, biomedical) is possible using the externally applied magnetic field as well by doping/mixing non-magnetic material (e.g., silica, graphene oxide, latex, polymer, gold) [16–22]. Similar way, micron-sized silica particles added magnetic fluid demonstrated an enhancement in the magnetic field induced self-assembly and magnetorheologi-cal properties [15,27,28]. Whereas, the colloidal silica nanoparticles mixed with the magnetic fluid, it results in a magnetic field induced high-yield stress (like magnetorheological fluid) [22]. The magnetic fluid prepared similarly also exhibited an increase in birefringence depending on the concentration of silica nanoparticles [23]. Despite this, the nature of the interaction of silica nanoparticles with the magnetite nanoparticles was not discussed.

Here, we report magnetite magnetic fluid mixed with colloidal silica nanoparticles. The suspension of colloidal silica nanoparticles (dia $\cong$ 12 nm) was simply added/mixed with the aqueous magnetic fluid. The system thus prepared is a physical mixture; without having any chemical, thermal treatment. The system reported here is different from the chemically synthesized $Fe_3O_4$–$SiO_2$ core-shell particles [9–12]. The magnetite magnetic fluid was synthesized using the chemical co-precipitation route. X-ray diffraction (XRD) data reveals a single-phase spinel structure. The interaction obtained from Fourier transform infrared spectroscopy and thermogravimetric analysis agree very well with microscopic data. The Scanning electron microscopic (SEM) images indicate the formation of nanorod like structure consists of magnetic nanoparticles covered with silica. Interestingly, superparamagnetic nature retains even after the addition of silica and in-spite of forming nanorod like structure. The magnetic field induced microscopy demonstrates micro-structure formations with controlled inter-chain spacing.

## 2. Experimental

### 2.1. Sample preparation

Analytical grade chemicals used were Iron (II) chloride (98%), Iron (III) chloride (97%), ammonia solution (28–30%), lauric acid ($\geq$98%), and LUDOX HS-40 colloidal silica (40 wt% suspension in $H_2O$) (all from Sigma-Aldrich (Merck)).

The synthesis of magnetic fluid has been carried out as follows [24,25]. Iron (II) chloride (6 g) and Iron(III) chloride (12 g), each dissolved in 25 ml water. The particles were precipitated by adding 30 ml $NH_4OH$ in iron chloride solutions at 300 K with vigorous stirring (450 RPM). After 5 min, luke-warm water (T~308 K) wash was given twice in order to remove the chlorine impurities, followed by 5% ammoniated water wash to provide charge to the freshly prepared particles. Ferrite slab magnets (surface magnetization 0.1 T) were used for the magnetic decantation process. Finally, thick magnetic slurry obtained was kept on a pre-heated hot plate with continuously stirring. Once the temperature of slurry reached around 353 K, 2.4 g lauric acid (LA) was added and gently stirred for 1–1.5 min. Thus double layers of lauric acid stabilized magnetite magnetic fluid (MF) formed. The fluid can be diluted many times (up to 50–60) due to the presence of extra surfactant. The fluid was dialyzed overnight (to remove extra free surfactant) and coded as F100 magnetic fluid (MF).

The colloidal suspension of silica nanoparticles (HS-40) (size ~ 12 nm) was stabilized using sodium counter ions. This suspension was mixed with the magnetic fluid to study the effect of silica nanoparticles on the magnetic fluid properties. Three samples were prepared by mixing X ml of F100 MF and Y ml of HS-40, whereas the total volume of the system kept constant (i.e., 1 ml). The first sample F80H20 consists of 0.8 ml F100 MF and 0.2 ml HS-40 suspension. The F75H25 contains 0.75 ml F100 MF and 0.25 ml HS-40 suspension. The F70H30 fluid has 0.7 ml F100 MF and 0.3 ml HS-40 suspension. It is to be noted that, the F80 fluid has 0.8 ml F100 MF, and 0.2 ml ammoniated water (5% $NH_4OH$ solution).

### 2.2. Characterization techniques

Powder X-ray diffractometer (XRD) D2-phaser (Bruker AXS, Germany) was used to analyze the crystal structure of $Fe_3O_4$ nanoparticles. The measurement was carried out from 10° to 70° in step of 0.02° with the scan rate of 5 s. Topas 4.2 software was used to analyze the XRD pattern.

Fourier Transform Infrared (FTIR) spectroscopic measurements were carried out using NICOLET 6700 spectrophotometer (Thermo). Each spectrum was recorded in the range of 400–4000 $cm^{-1}$ in an interval of 1 $cm^{-1}$ at 300 K. The sample pellet was prepared by thoroughly mixing the sample in KBr powder (IR grade, Sigma-Aldrich).

Thermogravimetric Analysis (TGA) (TGA/DSC-1, Mettler Toledo) was used to determine the thermal decomposition and phase transition temperatures. Hot air oven-dried (@ 100 $^{O}C$ - 16 h) samples (~15 mg) were placed in an alumina crucible and scanned in the temperature range of 50–1000 $^{O}C$ (under $N_2$ atmosphere). Initially, all the samples were scanned with 10 $^{O}C$/minute step size. After identifying the region of interest, the temperature range was segmented as follows: a) 50–200 $^{O}C$ -10 $^{O}C$/min.step size, b) 200–360 $^{O}C$ - 5 $^{O}C$/min. step size, c) 360–690 $^{O}C$ - 10 $^{O}C$/min. step size, d) 690–750 $^{O}C$ - 5 $^{O}C$/min. step size, e) 750–820 $^{O}C$ - 2 $^{O}C$/min. step size, and f) 820–1000 $^{O}C$ - 5 $^{O}C$/min. step size.

Magnetization measurements were carried out using vibrating sample magnetometer (7400, Lake Shore) at 300 K in the magnetic field range from 0 kA/m (0 T) to 1000 kA/m (1.2 T).

Microstructure formation was studied using an inverted metallurgical microscope (IM7100, Meiji Techno), equipped with a 20× objective lens (0.4 NA), and a ProgRes C3 CCD camera (Jenoptik). A drop of magnetic fluid, sandwiched between a glass plate and a coverslip, was subjected to a gradient magnetic field of 0.055T. Thus, the magnetic field applied was at right angles to fluid and incident light. ImageJ program was used to analyze microstructure formation.

## 3. Result and discussion

Fig. 1 shows x-ray diffraction (XRD) pattern (open symbol) of $Fe_3O_4$ nanoparticles. The parameters deduced from the Reitveld refinement (using TOPAS4.2 software) are composition: Inverse Spinel ferrite, single-phase spinel FCC structure, and space group: Fd3m. The whole pattern analysis carried out using fundamental parameter profile fitting (FPPF) is shown as a solid line in Fig. 1. The crystallite size determined using SPVII line shape profile is 11.2 $\pm$ 0.2 nm. Lattice parameter ($a$) obtained from the fit is 0.8388 $\pm$ 0.0001 nm, which agrees with the bulk (0.839 nm) [26].

The interaction of silica nanoparticles (HS-40) with the magnetite nanoparticle in the magnetic fluid stabilized using double layers of lauric acid (LA) surfactant is studied using the FTIR spectra. Fig. 2 shows FTIR spectra (% transmission) of F70H30, F75H25, F80H20 fluids along with its parent systems, i.e., F100 MF and silica suspension (HS-40). General observation of all spectra suggests the presence of all major peaks. The detailed analysis is given below.

In F100 MF, the IR band observed at 3440 $cm^{-1}$ is attributed to the stretching vibration of $H_2O$ molecules. Due to the lauric acid (LA) coating, H–C–H stretching vibrations are evident. The antisymmetric ($v_{as}$) and symmetric ($v_s$) modes of $CH_2$ seen at 2919 $cm^{-1}$ and 2850 $cm^{-1}$, respectively, whereas stretching of $CH_3$ are observed $v_{as}$ at 2955 $cm^{-1}$ and $v_s$ at 2870 $cm^{-1}$, respectively. The vibrational modes corresponding to carbonyl C=O chains usually observed in LA at 1712 $cm^{-1}$ disappeared here. Instead, two new vibrations corresponding to carboxylate asymmetric ($v_{as}COO^-$) and symmetric ($v_sCOO^-$) modes of vibrations appeared as broadband centered around 1635 and 1442 $cm^{-1}$, respectively ($\Delta = v_{as}COO^- - v_s COO^- \cong 193\ cm^{-1}$). The $\Delta$ depicts monodentate interaction between the carboxylate head and the magnetite nanoparticles of magnetic fluid. In other words, it confirms the chemi-adsorption of lauric acid on the surface of magnetite nanoparticles [27]. Herrera et al. reported similar interaction of magnetite nanoparticles with lauric acid and oleic acid [28]. The magnetite has an inverse spinel structure, where $Fe^{+2}$ occupies B site, while $Fe^{+3}$ occupies both A and B-sites. In the FTIR spectra, the band observed at 589 relates to the tetrahedral (A) site, whereas bands observed at 420 & 411 $cm^{-1}$ relates to the octahedral (B) site [29,30].

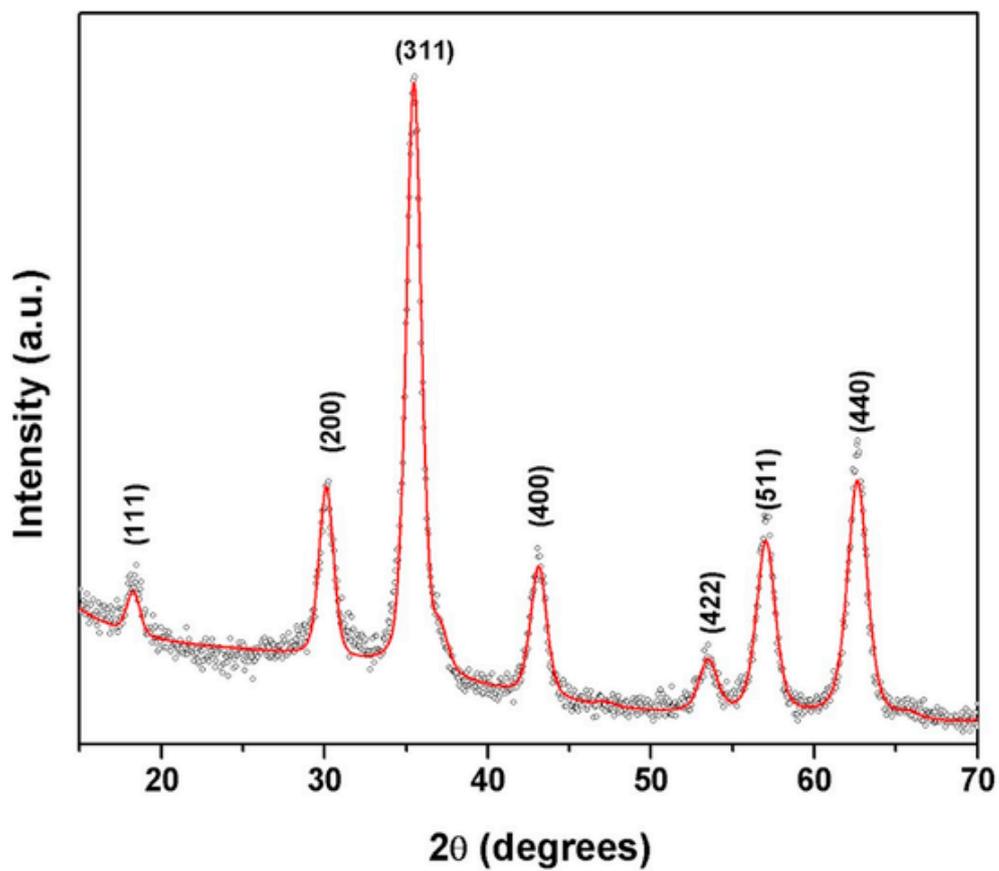

**Fig. 1.** X-ray diffraction data fitted using Rietveld refinement.

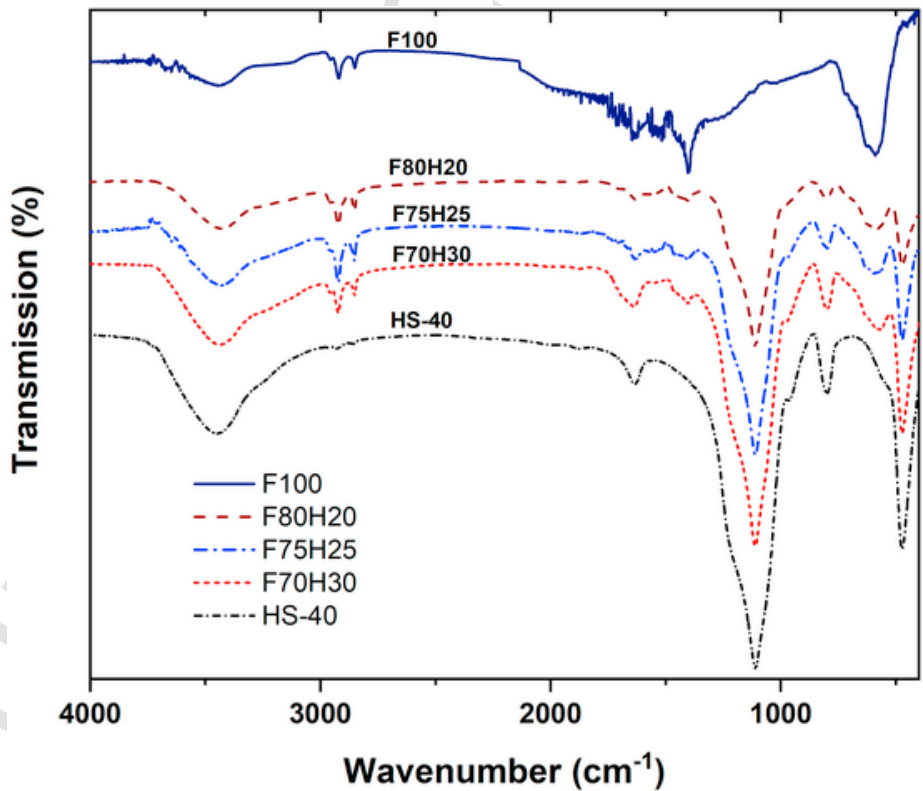

**Fig. 2.** FTIR spectra illustrating the influence of silica nanoparticles on F100, F80H20, F75H25, F70H30 magnetic fluids, and HS-40 silica suspension.

In the silica suspension (HS-40), the band observed at 3454 $cm^{-1}$ is attributed to the stretching vibration of $H_2O$ molecules. Correspondingly, the band at 1636 $cm^{-1}$ is due to the bending vibrations of $H_2O$ molecules. A shoulder around 3248 $cm^{-1}$ is due to the stretching vibrations of Si–OH (bonded water). The characteristics transverse optical (TO) frequency and longitudinal optical (LO) frequency bands of $SiO_2$ nanoparticles falls in the fingerprint region 700–1350 $cm^{-1}$ [31]. The TO and LO modes of the Si–O–Si asymmetric stretching vibrations are evident from the sharp band at 1111 $cm^{-1}$ with a shoulder around 1188 $cm^{-1}$, respectively. Correspondingly, band at 800 $cm^{-1}$ has a contribution from symmetric stretching vibrations of Si–O–Si and band at 471 $cm^{-1}$ assigned as O–Si–O bending vibrations. The observed bands agreed with those of synthesized amorphous $SiO_2$ [8].

The bands observed at 3439 $cm^{-1}$, 3435 $cm^{-1}$, and 3427 $cm^{-1}$, respectively, for F70H30, F75H25, and F80H20 are of $H_2O$ stretching vibration and indicate a redshift compared to both the parent fluids. Similarly, $H_2O$ bending vibrations are observed at 1639 $cm^{-1}$, 1635 $cm^{-1}$, and 1627 $cm^{-1}$, respectively, for F70H30, F75H25, and F80H20 fluids. The observed band varies compared to HS-40 silica suspension. The antisymmetric and symmetric stretching modes of $CH_2$ remains constant around 2922 $cm^{-1}$ and 2852 $cm^{-1}$ in all three fluids, which agrees with the parent F100 fluid. The TO and LO modes of the Si–O–Si asymmetric stretching vibrations remained constant at 1111 $cm^{-1}$, whereas variations in the shoulder position observed around 1214 $cm^{-1}$, 1221 $cm^{-1}$, and 1221 $cm^{-1}$ respectively for F70H30, F75H25 and F80H20 fluids. Corresponding symmetric stretching vibration of Si–O–Si at 801 $cm^{-1}$ and O–Si–O bending vibrations at 471 $cm^{-1}$ remained invariant for all the three fluids. Shoulder banding in HS-40 silica at 966 $cm^{-1}$ shifts to 973 $cm^{-1}$, 974 $cm^{-1}$, and 975 $cm^{-1}$, respectively, in F70H30, F75H25, and F80H20 fluids are associated with Si–O stretching vibration. The shift indicates the interaction of silica with the lauric acid coated magnetite. Also, A-site of Fe–O stretching vibration shifts from 589 $cm^{-1}$ to 576 $cm^{-1}$, 585 $cm^{-1}$, and 587 $cm^{-1}$, respectively, for F70H30, F75H25, and F80H20 fluids, while B-site Fe–O stretching vibration remains invariant. Here, we recall that the silica suspension was added to the ferrite magnetic fluid (without any chemical process); hence, it is unlikely to expect any change in the fingerprint region of ferrite particles (i.e., cation distribution band positions). However, as per the present spectral analysis, it can be concluded that silica NPs interacts with the lauric acid coated magnetite's. However, the observed variation in the cationic A-site of magnetite is surprising. In order to understand and further confirm the interaction of silica with F100 fluid, TGA experiments were carried out.

### 3.1. Thermogravimetry analysis

Thermogravimetric data are used to determine the decomposition temperature of the surfactant and phase-transition temperature of ferrite and silica nanoparticles. Fig. 3 shows the first derivative $\frac{dM}{dT}$ of F100, F70H30, F75H25, and F80H20 magnetic fluids. As discussed below, spectra show five peaks, corresponding to the transition temperature. The mass loss observed at ~120 °C is attributed to the absorbed moisture and/or water crystallization temperature. In F100 MF, transition temperatures 255 (±5) °C and 346 (±5) °C correspond to the decomposition of secondary and primary layers of chemi-adsorbed lauric acid present on the magnetite nanoparticles. The synthesis route adopted here lead to form double layers of surfactant on the magnetic nanoparticles [24,25]. A third peak corresponding to free/un-bound lauric acid surfactant is absent here. Hence, it confirms our analysis of FTIR spectra, where peak ~1712 $cm^{-1}$ corresponding to physi-adsorbed/unbound lauric acid peak is absent. Albeit, the surfactant remained only as a double layer of chemi-adsorbed lauric acid on the surface of magnetic nanoparticles. The addition of silica in the F70H30, F75H25, and F80H20 fluids changed the first peak from 255 (±5) °C (i.e., F100) to 262 (±5) °C, which is attributed to the interaction of silica nanoparticles with the second layer of LA. This result agrees with the interaction observed in the FTIR spectra. As expected, the first layer of LA did not alter on the addition of silica NP and remained constant, i.e.,~346 (±5) °C. Further, the increase in the temperature from 500 to 1000 °C illustrates the phase transition of magnetite. The peak ob-served in F100 at 700 ± 5 °C is attributed to the phase transition from FCC spinel ferrite $Fe_3O_4$ to wustite Fe–O [32,33]. Noticeably, we observed a redshift in the first phase transition from 700 (±5) to 733 (±5) °C, 734 (±5) °C and 724 (±5) °C in the F70H30, F75H25, and F80H20 respectively, and correlated with the reduction in the silica NP concentration. The secondary transition seen in F100 at 864 (±5) °C is due to further transformation to metallic Fe [34]. This peak moves to 799 (±2) °C, 795 (±2) °C, and 776 (±2) °C in respectively F70H30, F75H25, and F80H20 possibly reveals degradation of different structure. Perhaps during the heating cycle, Fe–C/Fe–Si or similar composite forms and may degrade [32].

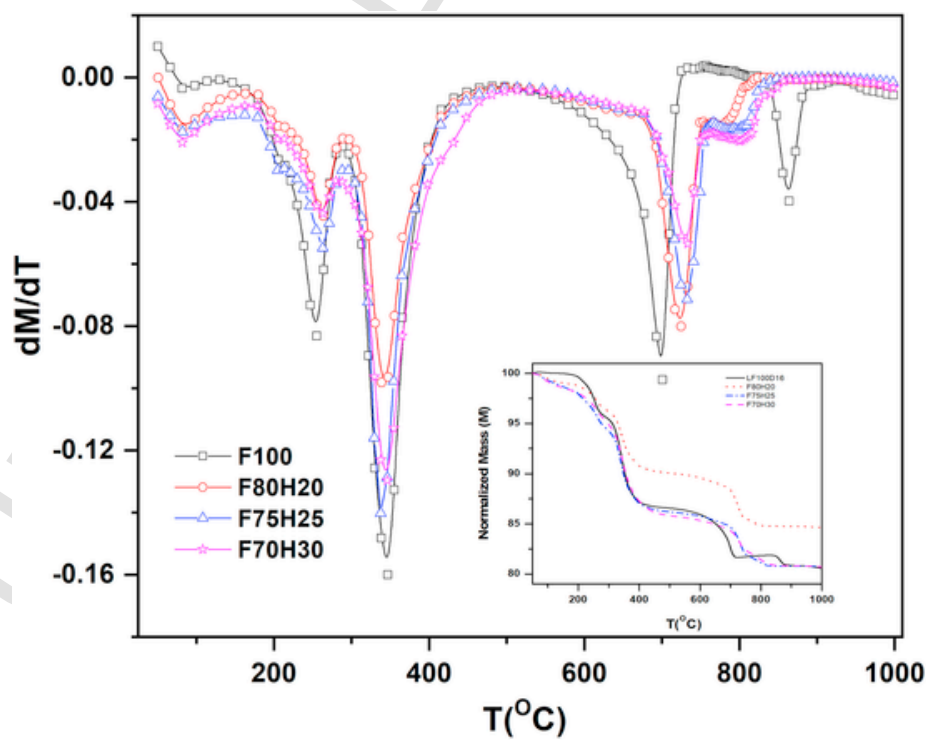

**Fig. 3.** First order derivative of thermal decomposition of F100, F80H20, F75H25, and F70H30. Inset: Variation in the mass loss (normalized) with increasing temperature.

The TGA analysis of F100, F70H30, F75H25, and F80H20 magnetic fluids used to calculate the number of organic ligands (molecules) bound with the magnetite surface using the relation [29,35], $p = \frac{m_2}{m_1+m_2} \times 100 \%$. Here, sample mass $m_1 = \rho_1 \times V_1$, with $\rho_1$ the density of bulk magnetite ($5.17 \times 10^3$ $kg/m^3$), the volume of each magnetite crystals $V_1 = \pi D^3/6$, with D the particle diameter of magnetite obtained from Scherrer's formula. The total mass of the organic molecules bound to one nanoparticle, $m_2 = n\overline{M_2} \times 10^{-3} \times 6.022 \times 10^{-23}$ $kg/mol$ measured as mass-loss during carbonization process, with $\overline{M_2}$ the molecular of lauric acid ($200.318 \times 10^{-3}$ $kg$). Thus, the number of ligands (molecules) bound at each mass-loss peak is calculated as,

$$n = \frac{\pi \times D^3 \times \rho_1 \times p \times 6.022 \times 10^{23}}{6 \times \overline{M_2}(100-p)} \quad (1)$$

Table 1 shows the number of ligands bound as the secondary and primary layers of lauric acid surfactant on a single magnetite nanoparticle derived from equation (1). In F100 fluid, the number of primary and the secondary molecules on single magnetite are 886 and 421 respectively, and hence the ratio of primary: secondary is 2.1. It infers that around each magnetite nanoparticle, there will be 2.1 lauric acid molecule in the first layer with corresponding one lauric acid molecule in the second layer. In F80H20 fluid, the number of lauric acid molecules in the primary and secondary layers are comparatively less than F100 fluid; however, the ratio of primary: secondary ligands increases. Whereas, in F75H25 and F70H30, the number of ligands and overall primary: secondary ratio increases to 2.79 and 3.21, respectively. The increase in this ratio may affect the energy ratio, namely magnetic and thermal energies. It may be noted here that the number of molecules in the primary and secondary layers is high in the F75H25 compared to the other two fluids (i.e., F70H30 and F80H20), which may affect on the size parameter of silica-magnetic cluster/aggregates.

Fig. 4 shows SEM images of (a) F80H20, (b) F75H25, and (c) F70H30 fluids. It emerged from the images that the addition of silica leads to form rod-type aggregates. The number density of this elongated/rod-shaped structure is higher in F80H20 fluid, while it relatively decreases with increasing silica concentrations. Also, the background becomes more whitish with increasing silica concentrations (i.e., a to c). General observation of all the images indicates formations of rod-like structure, but the presence of spherical and other shapes is evident. The average length of these rods are 1.43 μm, 1.49 μm, and 1.53 μm, while the average width 0.24 μm, 0.42 μm, and 0.64 μm respectively for F80H20, F75H25, and F70H30 fluids. Based on this, roughly, it can be estimated that the thickness of the rod-like/elongated structure increases with the increase in silica concentrations. This observation supports our TGA findings. Similarly, adding carboxymethylcellulose (CMC) *in-situ* and *ex-situ* in $Fe_3O_4$ nanoparticles leads to form two different types of structure [36].

A recent report discussed the preparation of magnetic silica microbeads using the water-in-oil emulsion method [37]. Scanning Transmission Electron Microscopy (STEM) resembles core-shell like structure, where magnetic nanoparticles layered inside silica microbeads. The surface porosity varies with span 80 (surfactant) concentrations. In the present work, magnetic fluid and silica suspension are mixed (unlike emulsion), which produces nanorods or rod-like structures. The white color spherical-like aggregates formed of either silica or silica-magnetite composites. To understand the magnetic nature of the observed nanorods, we performed magnetic measurements. Fig. 5 shows magnetization data of magnetic fluids F100, F80H20, and F30H70. We noted here that the saturation magnetization, initial susceptibility, mean magnetic size, and log-normal particle size distribution (σ) varies linearly with the magnetic volume fraction of F100 (see Table 2). Supplementary S1 discusses the analysis of magnetic properties. Nanorods possess uniaxial anisotropy and relatively high magnetic susceptibility compared to single-domain magnetic nanoparticles. In-spite of forming nanorod structure, we assume here that magnetite particles stuck in each rod behave like single-domain magnetite. In order to have a clear understanding of this behavior, we performed the magnetic field induced microscopy.

The microscopic images were captured from 0 to 5 min (300 s) with 30 s interval of applying 0.055T magnetic field for F80, F80H20, F75H25, and F70H30 fluids. Fig. 6 shows representative microscopic images from 1 to 5 min. The structure formed in the F80 is relatively scattered, whereas silica added fluids exhibit a relatively large and continuous structure. Moreover, the structure formed in the silica added fluids tends to stabilize over the period. Fig. 7(a–c) shows microstructured parameters, i.e., chain length, chain width, and successive dis-tance, determined using ImageJ software. It depicts that on the application of the magnetic field, in initial 30 s, field-induced structure formation takes shape. In the water-based magnetic fluid, the existence of pre-agglomerates helps to form a relatively stable field-induced structure [20,23]. Interest in understanding field-induced self-assembly in the superparamagnetic colloids in the presence of nano/micron non-magnetic particles also increases [38]. In the F80, F80H20, F75H25, and F70H30, the average length of the chains is 42.5, 47.5, 42.5, and 40 μm, while average width 4.75, 3.75, 4.00, and 3.5 μm respectively. Also, in all the fluids, the successive distance increases with time. Noted here that time-dependent variation in the successive distance is minimum in the F70H30 fluid, whereas maximum in the F80 fluid. This shows that the addition of silica inhibits the growth of the chains. Based on the data of average length and width of the nanorods (refer SEM data), one can estimate the numbers of nanorods to form field-induced chains. The average number of nanorods to form vertical chains are around 35, 29, and 26, respectively, in the F80H20, F75H25, and F70H30 fluids. Similarly, the number of nanorods estimated for the width of the chains is 16, 10, and 5, respectively, for the F80H20, F75H25, and F70H30 fluids. We propose the following hypothesis to explain the reduction in the chain dimensions while correlating with the FTIR and TGA analysis.

**Table 1**
The mass loss (%) and the decomposition temperature of F100, F80H20, F75H25, and F70H30 fluids. The number of the secondary and primary layers of lauric acid bound on a magnetite nanoparticles derived from equation (1), and increment in the ratio of primary: secondary with the increase in the silica nanoparticles are shown.

| Sample | Layer | T(°C) (±5) | % Wt loss dM/dT | No. of Ligands | The ratio of primary to secondary |
|---|---|---|---|---|---|
| **F100** | Secondary | 255 | 8.3 | 421 | 2.10 |
|  | Primary | 346 | 16 | 886 |  |
| **F80H20** | Secondary | 262 | 4.4 | 214 | 2.35 |
|  | Primary | 346 | 9.8 | 505 |  |
| **F75H25** | Secondary | 262 | 5.5 | 271 | 2.79 |
|  | Primary | 346 | 14 | 757 |  |
| **F70H30** | Secondary | 262 | 4.4 | 214 | 3.21 |
|  | Primary | 346 | 12.9 | 689 |  |

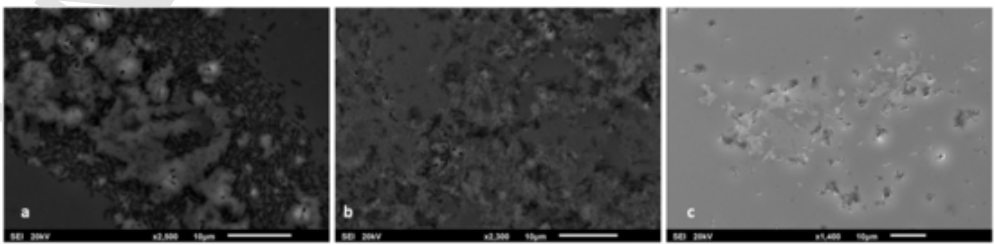

**Fig. 4.** SEM images illustrating the formation of nanorods in (a) F80H20, (b) F75H25, and (c) F70H30 fluids.

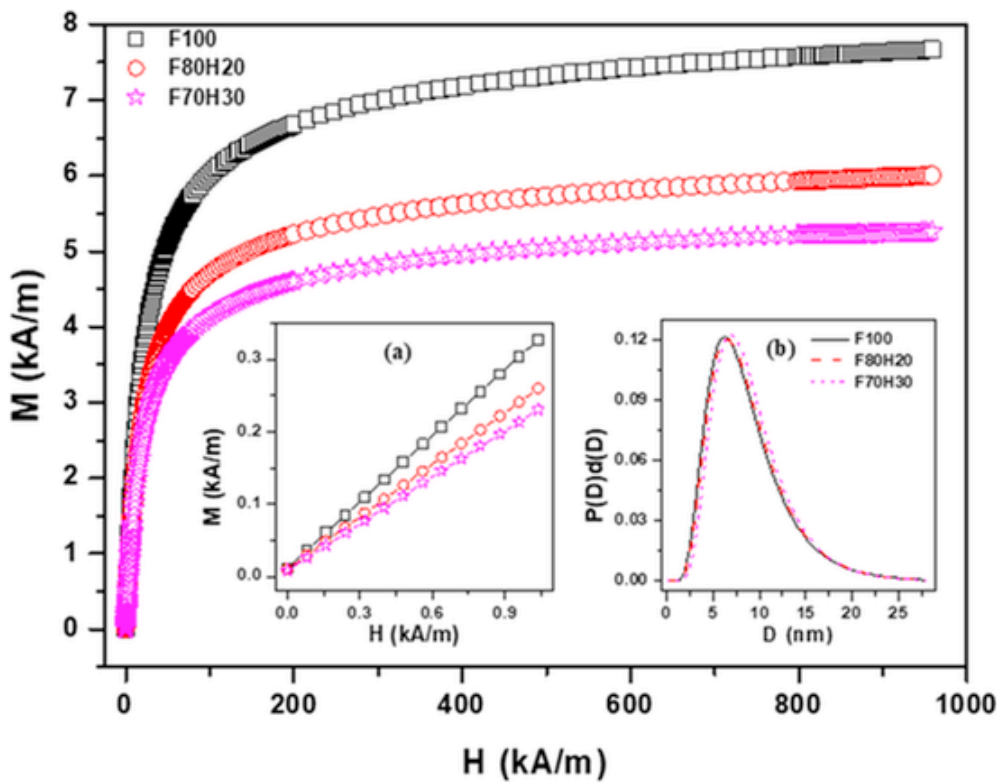

**Fig. 5.** Magnetic field-induced variation in magnetization at 300 K of F100 and the corresponding silica added samples F80H20 and F70H30. Inset figures (a) initial susceptibility, and (b) log-normal size distribution of all the three samples.

**Table 2**
Parameters derived from the data are initial susceptibility ($\chi_i$), saturation magnetization ($M_s$), and magnetic volume fraction ($\phi_m$) (assuming $M_d$ constant). Other parameters derived are mean particle diameter ($D_m$), volume-weighted mean magnetic diameter ($D_{mV}$), number weighted mean diameter ($D_{mN}$) & log-normal size distribution ($\sigma_D$). (refer supplementary S1 for detail).

| Sample | $\chi_i$ (±0.001) | $M_s$ (kA/m) (±0.0045) | $\phi_m$ | $D_m$ (nm) (±0.1) | $D_{mV}$ (nm) (±0.01) | $\sigma_D$ |
|---|---|---|---|---|---|---|
| F100 | 0.305 | 8.1843 | 0.00169 | 11.3 | 8.09 | 0.47 |
| F80H20 | 0.241 | 6.3953 | 0.00132 | 11.4 | 8.29 | 0.46 |
| F70H30 | 0.214 | 5.5224 | 0.00114 | 11.5 | 8.64 | 0.44 |

The lauric acid (surfactant) and silica nanoparticles both are non-magnetic in nature. The silica nanoparticles used here are stabilized using sodium ions. In other words, sodium counter ions prevent silica from aggregation. On adding silica suspension in the magnetic fluid, it is reacting with the outer (second) layer of lauric acid. The double-layer of surfactant interacting with silica nanoparticles should influence on the short-range (van-der Waals and steric repulsion) and long-range (dipole-dipole) interactions. The interaction energy of magnetic fluid comprises of magnetic and non-magnetic particles (here silica) is represented by [39],

$$U_{ij} = -\frac{\mu_f m_1 m_2}{4\pi r_{ij}^3} \left[1 - 3\cos^2\theta\right] \quad (2)$$

where, $m_1 = \left|V_{p1}\frac{3(\chi_{p1}-\chi_f)}{(\chi_{p1}-\chi_f)+3(\chi_f+1)}\boldsymbol{H}_a\right|$, and $m_2 = \left|V_{p2}\frac{-3\chi_f}{2\chi_f+3}\boldsymbol{H}_a\right|$ represents the absolute values of the equivalent magnetization for the magnetic and non-magnetic particles, respectively. The magnetic interac-

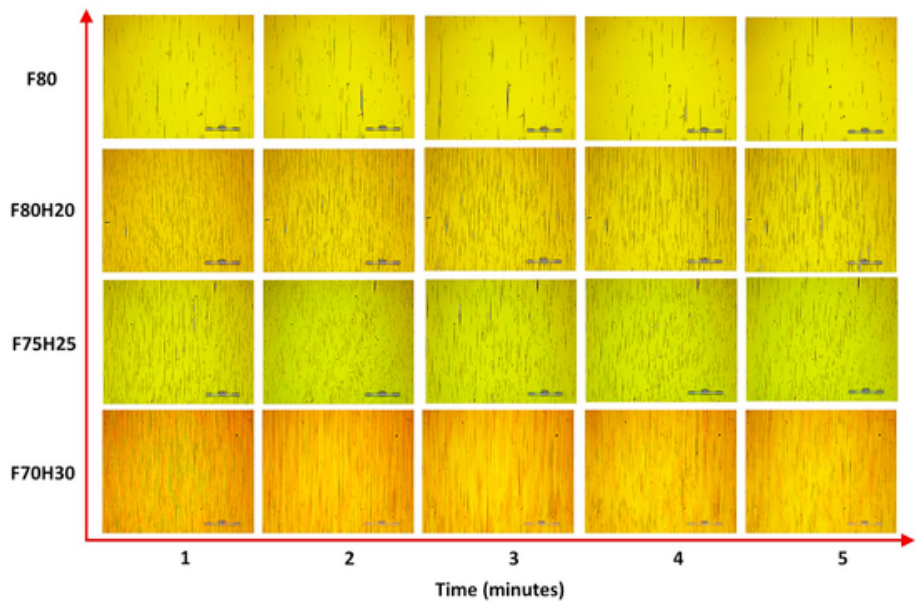

**Fig. 6.** Microscopic images of F80, F80H20, F75H25, and F70H30 fluids at different times after applying the 0.055T magnetic field. Scale 400 μm.

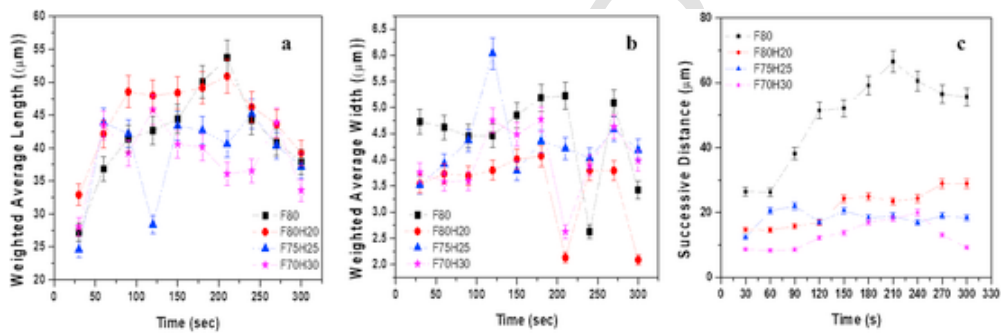

**Fig. 7.** Time-dependent weighted average (a) length, (b) width of the chains, (c) successive interchain distance of the magnetic field induced microstructure.

tion force is given by [39],

$$F_{ij} = -\nabla U_{ij} = -\frac{3\mu_f m_1 m_2}{4\pi r_{ij}^4}\left[(1-3\cos^2\theta)e_r + 2\sin\theta\cos\theta\, e_\theta\right] \quad (3)$$

Recall here that $\theta$ represents the angle between the linking line of the two particles and the direction of the applied magnetic field, $e_r$ the coordinate variable along the linking line of the two particles, and $e_\theta$ the cylindrical coordinate system. The trends of the interactive energy and the magnetic force vary with the angle $\theta$, and both the magnetic and non-magnetic particles exhibit the opposite direction of the magnetization (refer equ. 2 & 3). The particle interaction is repulsive when $0° \leq \theta < 54.73°$ [40], and attractive when $54.73° < \theta \leq 90°$ [39]. Which is contrary to the magnetic fluid comprises of only magnetic particles, which shows that the interparticle interaction is positive when $0° \leq \theta < 54.73°$. The effect of magnetic and non-magnetic particles are discussed in detail in Ref. [39].

In the present fluids, we approximated numbers of particles/aggregates in a chain based on the nanorod dimensions derived using SEM. In the F80H20, F75H25, and F70H30 fluids, the average nanorods assembled are longitudinal (and transverse) 34 (16), 29 (10), and 26 (7), respectively. As it is observed that in all three fluids, magnetic field induced structure stabilizes after 1 min (60 s), numbers reported here are after 1 min only. As shown in Fig. 2 of Ref. [39], the magnetic and non-magnetic particles parallel to the field lines experience a repulsive force, whereas perpendicular to the field lines experiences an attrac-tive force. Similar is reflected in the present fluids, where higher silica concentration leads to form short and thin chains, whereas low silica concentration forms long and thick chains. Fig. 7(c) shows the successive interchain distance for all the fluids. In the magnetic fluid F80 large change is observed with the increase in time, while in silica added fluids the interchain distance decreases relatively. The estimated number of lines in a grating per meter ($N = l/d$) (l = length of a grating (say 1 m), and d = interchain distance (in meter)) is in the range of 150–350, 350–670, 450–600, and 550–1150 respectively for the F80, F82H20, F75H25 and F70H30 fluids. Although we need to achieve better control on these deviations, but resultant range of F75H25 is relatively less. The diffraction angle ($\theta_r$) is calculated as, $\sin\theta_r = m\lambda/d$, where m = order to diffraction, and $\lambda$ = wavelength of incident light (here 650 nm). It infers that in case of first order diffraction (m = ±1), $\theta_r$ ranges from 0.56° to 1.41°, 1.3°–2.5°, 2.0°–3.0°, and 1.87–4.3° respectively for the F80, F82H20, F75H25 and F70H30 fluids. These preliminary results are encouraging. Further, field dependent experiments will throw some more light to develop tunable diffraction grating.

## 4. Conclusion

Indirect interaction of nonmagnetic particles with the magnetic nanoparticles in-terms of optical, magnetic, rheology, etc. properties are reported in the literature [41,42]. However, shreds of evidence of direct interaction and its influence on various properties are scare [37]. Here, we demonstrate addition of silica nanoparticles suspension (Ludox HS-40) to the aqueous magnetic fluid (magnetite nanoparticles sta-bilized using double layers of lauric acid surfactant) results in chemically interactive fluid. The chemical interaction is explained using FTIR and TGA. The analysis of the FTIR spectrum suggests that silica nanoparticles interact with the lauric acid through Si–O, and the intensity of its band increases with increasing silica concentrations. The interaction of silica nanoparticles with lauric acid is further confirmed using TGA, which reveals that silica bind with the outer (second) layer of lauric acid attached to the magnetite nanoparticles. The ratio of primary to secondary layer ligands increases with increasing silica concentration. However, the number of lauric acid ligands observed is relatively more in F75H25 fluid compared other two fluids containing high and low silica concentration. It indicates that surfactant layer thickness is subjected to the concentration of nonmagnetic silica nanoparticles. The scanning electron microscopy indicates the formation of silica decorated magnetite nanorods. Despite of forming nanorod structure, the magnetic fluid exhibits superparamagnetic behaviour. The magnetic field assisted assembly indicates the formation of long, thick and scattered chains in the aqueous magnetic fluid. Whereas, relatively short, dense chains with controlled interchain distances are observable in the silica added magnetic fluids. This chains behaves like diffraction grating lines. The preliminary results of grating parameters are encouraging and has potentiality to develop as tunable diffraction grating.

## CRediT authorship contribution statement

**Nidhi Ruparelia:** Formal analysis. **Urveshkumar Soni:** , Formal analysis. **Rucha P. Desai:** Formal analysis, Supervision. **Arabinda Ray:** Formal analysis.

## Declaration of competing interest

The authors declare that they have no known competing financial interests or personal relationships that could have appeared to influence the work reported in this paper.

## Acknowledgments

The work is carried out under EMR/2016/002278 project sponsored by the Science and Engineering Research Board (SERB) , Department of Science and Technology (DST), India.

## Appendix A. Supplementary data

Supplementary data to this article can be found online at https://doi.org/10.1016/j.jssc.2020.121574.